\documentclass[aps,prb,twocolumn]{revtex4-1}
\usepackage{color}
\usepackage{graphicx}
\usepackage{cancel}
\usepackage{hyperref}
\usepackage{amsmath}
\usepackage{amsfonts}
\usepackage{ulem}
\newcommand{\stkout}[1]{\ifmmode\text{\sout{\ensuremath{#1}}}\else\sout{#1}\fi}

\def  \bsig    {\mbox{\boldmath$\sigma$}}
\def  \bgam    {\mbox{\boldmath$\Gamma$}}

\usepackage[utf8]{inputenc}
\usepackage{t1enc}

\usepackage{xcolor}

\begin{document}

\title{Conduction of surface electrons in a topological insulator \\
with spatially random magnetization}

\author{S. Kud{\l }a$^{1}$, A. Dyrda{\l }$^{2,3}$, V. K. Dugaev$^{1}$, J. Berakdar$^{2}$ and J. Barnaś$^{3,4}$}
\affiliation{$^1$Department of Physics and Medical Engineering, Rzesz\'ow University of Technology,
Al.~Powsta\'nc\'ow Warszawy 6, 35-959 Rzesz\'ow, Poland\\
$^{2}$Institute of Physics, Martin-Luther Universit\"at Halle-Wittenberg, 06099 Halle (Saale), Germany\\
$^{3}$Faculty of Physics, Adam Mickiewicz University in Pozna\'n, ul. Uniwersytetu Pozna\'nskiego 2, 61-614 Pozna\'n, Poland\\
$^{4}$Institute of Molecular Physics, Polish Academy of Sciences, 60-179 Pozna\'n, Poland}

\begin{abstract}
Using the Green functions method we study  transport properties of surface electrons in topological insulators in the presence of  a correlated random exchange field. Such an exchange field may be due to random magnetization with correlated fluctuations. We determine the relaxation time due to scattering from the magnetization fluctuations and from other structural defects. Then we calculate the longitudinal charge conductivity taking into account the contribution due to vertex correction.
\end{abstract}
\date{\today }
\maketitle


{\it Introduction:}
Topological properties of matter are currently at the frontline of research in condensed matter physics~\cite{Hasan_2010,Qi_2011}. Much attention has been focused recently on surface states in topological insulators (TIs), where electrons at the surface behave as massless Dirac fermions with spin-momentum locking~\cite{Xia_2009,Hsieh2009,Niu_Shen_2010}. This locking is responsible for many spin-dependent transport phenomena~\cite{Konig_2007,Xu_2016,Akzyanov_2018,Akzyanov_2019,He_Vignale2018,Lv_2018,Fert2016}. It is known, that owing to the proximity-induced exchange interaction, a thin ferromagnetic layer deposited on the surface  of a TI opens a gap at the Dirac point in the  electronic spectrum~\cite{Zhang_2009,Kou_2013,Fan_2014}.
Generally, hybrid systems based on TIs and thin ferromagnetic films, or on TIs with  surfaces intentionally decorated by magnetic adatoms bear great potential for spin-to-charge conversion phenomena~\cite{Zhang_2012,Xu_2014,Morimoto_2015}.
Proximity-induced exchange field not only affects the spectrum of  electronic surface states but also  introduces unavoidable magnetic disorder at the interface, which influences the relaxation processes of surface electrons in TIs.

As well established,  magnetic disorder has a strong impact on the  transport properties~\cite{Wang_2016}.  Such disorder may have different sources, for instance, the intrinsic magnetic properties of amorphous alloys, such as local magnetization, magnetocrystalline and exchange energies, etc, fluctuate in space due to a random distribution of magnetic ions (as a consequence of external and internal stresses)~\cite{Kronmuller_1978,Bernal_1959,Bernal_1960}. Spatially random magnetic fields may also be realized in semiconductor heterostructures, where they lead to interesting magnetotransport phenomena, including negative magnetoresistance due to weak localization, positive magnetoresistance related to small angle scattering of ballistic electrons by flux tubes or the presence of extended states analogous to quantum Hall edge states in random magnetic field with zero average value. Random magnetic fields in 2DEG heterostructures are realized, e.g., by capping the sample with superconducting film that ensures inhomogeneous distribution of magnetic flux tubes when an external magnetic field is applied (see, e.g., references~\onlinecite{Bending_1990,Bending_1990_prl,Geim_1992,Geim_1994}). Another possibility is to place the rough demagnetized permanent magnet (such as NdFeB) on top of the  heterostructure surface with two-dimensional electron gas~\cite{Mancoff_1995,Mancoff_1996}.

Thus, one may also expect a number of transport phenomena induced by magnetic disorder at the surface of TIs. For instance,
it was shown that the presence of magnetic impurities at the surface of TIs may lead to the in-plane magnetoresistance which is a mixture of the anisotropic and spin magnetoresistance,  as well as to planar Hall effect~\cite{Bauer}.  Moreover, random magnetic impurities may lead to opening an energy gap for the edge states~\cite{Erlingsson,Ernst,Piper}, and also may improve quality of the quantum anomalous Hall effect in magnetic TIs~\cite{Xing}. Recently the electron states at the surface of a TI attached to a ferromagnet, described by the  \textit{XY} model, have also been considered, and it was shown that the classical magnetic fluctuations in the ferromagnet could be mapped onto the problem of Dirac fermions in the random magnetic field~\cite{Galitski}.
Here, we consider   a model system where the spatially fluctuating magnetization interacts with the surface electrons in TIs due to an exchange field. We assume that the average value of magnetization (or exchange field) vanishes, However, $\langle M({\bf r})\, M({\bf r'})\rangle$, is finite and   is described by a given correlation function $\langle M({\bf r})\, M({\bf r'})\rangle = \mathcal{C}(\mathbf{r} - \mathbf{r}')$. We show that such fluctuations have a significant impact on the transport properties of surface electrons.


{\it Model:} We consider  2D electrons at the surface of a topological insulator  in a random magnetization field (equivalently random exchange field). The system is described by the following single particle Hamiltonian:
\begin{eqnarray}
\label{H}
\hat{H}=-iv\bsig \cdot \nabla +M({\bf r})\, \sigma _z\, ,
\end{eqnarray}
where $v = \hbar v_{F}$ and  $v_{F}$ is the electron (Fermi) velocity, $M(\mathbf{r})$ is a random magnetization (exchange) field measured in energy units, and ${\bsig = (\sigma_{x}, \sigma_{y}, \sigma_{z})}$ is the vector of the Pauli matrices operating in the spin space. Here we assume that $M({\bf r})$ vanishes on average,
 $\langle M({\bf r})\rangle =0$, but the second statistical moment is finite and is given in the form  $\langle M({\bf r})\, M({\bf r'})\rangle =  \mathcal{C}(\mathbf{r} - \mathbf{r}') = \langle M^2 \rangle\, g(|{\bf r}-{\bf r'}|)$. Moreover, we also assume that all higher even-order statistical moments are reduced  to the second-order one, whereas the odd-order correlators vanish. The correlation function $g(|\mathbf{r} - \mathbf{r}'|)$  carries information on the characteristic correlation length  $\xi$ of the fluctuations. The Fourier transform of $\mathcal{C}(\mathbf{r} - \mathbf{r}') $ has the Gaussian form: ${C(\mathbf{q}) = \langle M^{2} \rangle \xi^{2} \exp^{-q^{2}\xi^{2}}}$. 

 In the following calculations, the influence of the random magnetization on the  transport properties is treated perturbatively. The Green function corresponding to the unperturbed  Hamiltonian reads
\begin{eqnarray}
\label{G_R}
G^{0R}(\varepsilon ,{\bf k})=\frac{\varepsilon +v\bsig \cdot {\bf k}}
{(\varepsilon -\varepsilon_{1 k}+i\delta )(\varepsilon -\varepsilon_{2 k}+i\delta )},
\end{eqnarray}
where $\varepsilon _{1,2\, k}=\pm vk \equiv \pm \varepsilon_{k}$ are the eigenvalues of the unperturbed part of the Hamiltonian (\ref{H}).
In the following   we show explicitly how the random magnetization affects the  relaxation time and the conductivity of the surface electrons in TIs.


{\it Relaxation time:}
With the Green function (\ref{G_R}), one can write the electron self energy due to scattering from the fluctuating exchange field in the Born approximation as
\begin{eqnarray}
\label{SE}
\Sigma ^R(\varepsilon ,{\bf k})= \int \frac{d^2{\bf k'}}{(2\pi )^2}\, \mathcal{C}(|{\bf k-k'}|)\,
\sigma _z\, G^{0R}(\varepsilon ,{\bf k'})\, \sigma _z \nonumber\\
= \int \frac{d^2{\bf k'}}{(2\pi )^2}\, \mathcal{C}(|{\bf k-k'}|)\,
\frac{\varepsilon -v\bsig \cdot {\bf k'}}
{(\varepsilon -\varepsilon _{k'}+i\delta )(\varepsilon +\varepsilon _{k'}+i\delta )}.
\label{5}
\end{eqnarray}

As we are interested in the relaxation time of quasiparticles  we focus on the  imaginary part of the self-energy, $\Im[\Sigma_{R}]$ only, i.e. we study
\begin{eqnarray}
\label{6}
\Sigma ^R(\varepsilon ,{\bf k})=- i \frac{ \pi }{2\varepsilon }\int \frac{d^2{\bf k'}}{(2\pi )^2}\, \mathcal{C}(|{\bf k-k'}|)\,
\big( \varepsilon -v\bsig \cdot {\bf k'}\big) \nonumber \\
\times\big[ \delta (\varepsilon -\varepsilon _{k'})+\delta (\varepsilon +\varepsilon _{k'})\big]
\equiv -  i\Gamma_{0} \sigma_{0} - i \bgam \cdot \bsig , \hspace{0.3cm}
\end{eqnarray}
where for convenience the following notation has been introduced:
\begin{eqnarray}
\label{8}
&&\Gamma_0= \frac{\pi}{2} \int \frac{d^2{\bf k'}}{(2\pi )^2}\, \mathcal{C}(|{\bf k-k'}|)\,
\big[ \delta (\varepsilon -\varepsilon _{k'})+\delta (\varepsilon +\varepsilon _{k'})\big],
\\
&&\bgam= \frac{-\pi}{2 \varepsilon}\int \frac{d^2{\bf k'}}{(2\pi )^2}\, \mathcal{C}(|{\bf k-k'}|)\, v {\bf k'}\,
\big[ \delta (\varepsilon -\varepsilon _{k'})+\delta (\varepsilon +\varepsilon _{k'})\big].\hspace{0.3cm}
\end{eqnarray}
Upon integrating over the wavevector $\mathbf{k}'$ one finds at the mass surface,  $k = k_{1} = |\varepsilon|/v$,  the solution
\begin{eqnarray}
\Gamma_{0} = \frac{ k_{1}}{4 v} \langle M^{2} \rangle \xi^{2} \exp(- \xi^2 2 k_{1}^{2}) \mathcal{I}_{0}(2 \xi^{2} k_{1}^{2}),
\\
\bgam = - \frac{\mathbf{k}}{k}\frac{\langle M^{2} \rangle}{4 \varepsilon} \xi^{2} k_{1}^{2} \exp(- \xi^{2} 2 k_{1}^{2}) \mathcal{I}_{1}(2 \xi^{2} k_{1}^{2}).
\end{eqnarray}
Note that $\mathcal{I}_{0}$ and $\mathcal{I}_{1}$ are the modified Bessel functions of the 0-th and 1-st kind, respectively.
Accordingly, the  averaged Green function in the weak scattering approximation has the form
\begin{equation}
\label{GRwithSE}
G^{R/A}(\varepsilon, \mathbf{k}) = \frac{\varepsilon \sigma_{0} + v \mathbf{k} \cdot \bsig}{(\varepsilon - \varepsilon_{k} \pm i \gamma_{1})(\varepsilon + \varepsilon_{k} \pm i \gamma_{2})},
\end{equation}
where
\begin{equation}
\gamma_{1,2} = \Gamma_{0} \pm \frac{\mathbf{k}}{k}\cdot\bgam + \gamma_{0}.
\end{equation}
Here we have added  $\gamma_{0}$, that stands for scattering rate on all other defects in the system --- $\gamma_{1,2} \rightarrow \gamma_{0}$ when $\langle M^{2}\rangle \rightarrow 0$.
Thus, the relaxation rate can be cast as
\begin{eqnarray}
\gamma_{1,2} = \frac{k_{1}}{4 v} \langle M^{2} \rangle \xi^{2} \exp(-2\xi^{2} k_{1}^{2})\hspace{2.5cm} \nonumber\\
\times \left[ \mathcal{I}_{0}(2\xi^{2} k_{1}^{2}) \mp {\mathrm{sign}}(\varepsilon) \mathcal{I}_{1}(2\xi^{2} k_{1}^{2})\right] + \gamma_{0}.
\end{eqnarray}%
 Note that the expression for relaxation rate is the same for the negative and positive energy branches, $\gamma_{1} = \gamma_{2} \equiv \gamma = \gamma_M +\gamma_0$, where $\gamma_M$  is fully determined by the intrinsic properties of the topological insulator, i.e., by $v_{F}$, $k_{F}$, and two parameters describing the spatial magnetization fluctuations, that is $\langle M^{2} \rangle$ and $\xi$.
The electron relaxation time is given by the relation
\begin{equation}
\tau = \frac{\hbar}{2 \gamma},
\end{equation}
while the electron mean free path, $\Lambda_{\rm mf}$, reads
$\Lambda_{\rm mf} = v_{F} \tau = v/(2 \gamma)$.

\begin{figure}[t]
\centering
\includegraphics[width=1.0\columnwidth]{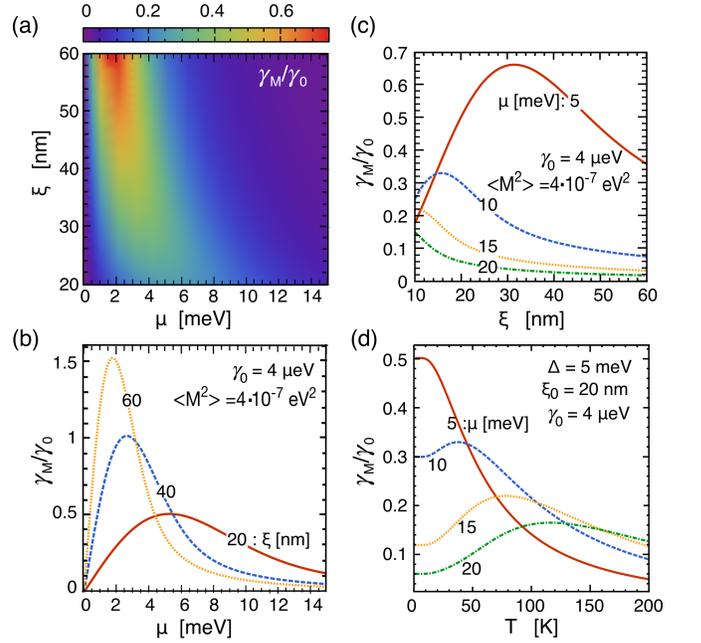}
\caption{Relaxation rate $\gamma_M$ due to scattering on the magnetization fluctuations, normalized to the relaxation rate $\gamma_0$ due to scattering on other defects. The amplitude $<M^2>$ of the fluctuations and $\gamma_0$ are assumed to be  constant, as indicated. Relative relaxation rate $\gamma_M/\gamma_0$ as a function of the Fermi energy $\mu$ and the correlation length $\xi$ (a) and the corresponding cross-sections for constant values of $\xi$ (b) and constant values of $\mu$ (c). Figures (a-c) correspond to zero temperature, while the temperature dependence of the relaxation rate $\gamma_M/\gamma_0$ is shown in (d). The Fermi velocity $v_{F} = 3.8\cdot10^{5}$m/s. Other parameters as indicated.}
\label{fig1}
\end{figure}

From the above formulas  one can immediately see how the random magnetization fluctuations influence the relaxation processes. Figure~\ref{fig1} presents the relaxation rate due to magnetic fluctuations, $\gamma_M$, calculated in the zero temperature limit and normalized to the relaxation rate $\gamma_{0}$ (i.e., to the relaxation  rate in the absence of magnetic fluctuations). In this figure $\gamma_0$ is considered as a parameter and is assumed constant and equal to $4\mu$eV.  Figure~\ref{fig1}(a) shows $\gamma_M/\gamma_0$  as a function of the correlation length $\xi$ and Fermi energy $\mu$. The range of parameters $\xi$ and $\mu$, where the scattering rate $\gamma_M$ is large is clearly seen in this figure. This behavior follows from the two factors in the correlation function: $\xi^{2}$ and $\exp ({-q^{2}\xi^{2}})$. The former factor reduces the electron scattering rate  for small values of $\xi$, while the second one reduces the scattering rate for large values of $\xi$, $\xi\gg(1/q)$, i.e. for the correlation length much longer than the electron wavelength $\lambda$ (the latter is determined by the Fermi energy $\mu$). Between these two limiting situations, the relaxation rate $\gamma_M$ acquires a maximum value, as clearly visible in Fig.~\ref{fig1}(a). This behavior can be also clearly seen in the corresponding cross-sections of Fig.~\ref{fig1}(a) for constant values of $\mu$ and $\xi$, as shown in Fig.~\ref{fig1}(b,c).

Interestingly, in the case of spatial fluctuations of magnetization one can assume that the correlation length may be a function of temperature, $\xi = \xi(T)$. Such a dependence is a consequence of the fact that the fluctuations of magnetization (fluctuations of magnetic moments of adatoms) depend in general on the temperature. In such a case, one can introduce the temperature dependence of the correlation length by the following phenomenological formula:
\begin{equation}
{\xi = \xi_{0} \left( 1 - \exp\left(-\frac{\Delta}{k_{B} T}\right)\right)},
\end{equation}
where $\xi_{0} = \xi(T=0)$ and $\Delta$ is the energy scale for magnetic interaction between the magnetic impurities.

Figure~\ref{fig1} (d) presents the relaxation rate $\gamma_M$  when  the correlation length decreases with increasing temperature according to the above equation.
For the assumed zero-temperature value of the correlation length, the fluctuations for a fixed chemical potential change character with increasing temperature from the long-range ($\xi\gg\lambda$) to short-range ($\xi\ll\lambda $) ones. Accordingly, a pronounced maximum at a certain temperature appears between these two regimes. Exception from this is the case of the lowest value of $\mu$ in Fig.~\ref{fig1} (d), where the low-temperature value of $\xi$ is comparable to the wavelength $\lambda$ so the corresponding curve starts from the point around the corresponding maximum.


{\it Vertex function and conductivity:} To determine the electrical conductivity, the vertex function is required. Thus,
 we write the dc  current density within the Kubo formalism~\cite{mahan} as
\begin{equation}
\label{59}
j_x=-\frac{evE_x}{2\pi}\, {\rm Tr}\int \frac{d^2{\bf k}}{(2\pi )^2}
\int d\varepsilon \, f'(\varepsilon )
J_x({\bf k})\, G^R_{\bf k}(\varepsilon )\, \sigma _x\, G^A_{\bf k}(\varepsilon ),
\end{equation}
 where $f(\varepsilon)$ is the Fermi-Dirac distribution function,  and $J_{x}$ is the renormalized current density vertex function. Assuming the mean free path is always longer than the correlation length of the magnetic fluctuations, $\Lambda_{\rm mf} \gg \xi$, the leading contribution to the vertex function originates from the  magnetization fluctuations (in other words, we neglect here e.g. non-magnetic scalar impurity potential and its contribution to the vertex correction).

The renormalized current vertex in the ladder approximation can be calculated from the self-consistent equation
{\small{
\begin{eqnarray}
\label{28}
{\bf J}({\bf k})=\hat{{\bf j}} + \int \frac{d^2{\bf k'}}{(2\pi )^2}  \mathcal{C}({\bf k-k'})  \sigma _z\,
G^A(\varepsilon ,{\bf k'})\, {\bf J}({\bf k'})\, G^R(\varepsilon ,{\bf k'}) \sigma _z ,\hspace{0.5cm}
\end{eqnarray}
}}
where $\hat{{\bf j}}=ev\bsig/\hbar$
is the operator of electrical current density. Generally, the current density vertex function  can be written in the form:
\begin{eqnarray}
\label{31}
{\bf J}({\bf k})=\frac{ev}{\hbar }\, \big( {\bf k}\, g_{0{\bf k}} \sigma_{0}+ g_{1{\bf k}} \bsig\big).
\end{eqnarray}
Inserting then Eq.(\ref{31}) into Eq.(\ref{28}) one finds the solutions for $g_{0{\bf k}}$ and $ g_{1{\bf k}} $ in the following form:
\begin{eqnarray}
\label{56}
&&g_{0k_1}=\frac{v}{\kappa} \frac{\eta }{\varepsilon} \mathcal{I}_{1}(2\xi^{2} k_{1}^{2}),
\\
&&g_{1k_1}= \frac{1}{\kappa} \left( 1-\eta \mathcal{I}_{1} (2\xi^{2} k_{1}^{2})\right),
\end{eqnarray}
where $\kappa$ and $\eta$ are defined as,
\begin{eqnarray}
\kappa = 1 - \eta \mathcal{I}_{1}( 2\xi^{2} k_{1}^{2}) + \eta \mathcal{I}_{2}( 2\xi^{2} k_{1}^{2}) + \frac{3}{2} \eta \mathcal{I}_{0}( 2\xi^{2} k_{1}^{2}) \hspace{0.2cm}
\nonumber\\ + \eta^{2} \mathcal{I}_{1}( 2\xi^{2} k_{1}^{2}) [\mathcal{I}_{0}( 2\xi^{2} k_{1}^{2}) + \mathcal{I}_{2}( 2\xi^{2} k_{1}^{2})],\hspace{0.2cm}\\
\label{58}
\eta =\langle M^{2} \rangle \xi^{2} \frac{k_{1}}{4 v \gamma} e^{- 2\xi^{2} k_{1}^{2}}.\hspace{4cm}
\end{eqnarray}
Importantly, the renormalized vertex function contains two components. One component (the second term in Eq.(\ref{31})) is in fact a simple renormalization of the current density operator, $\hat{\mathbf{j}} \rightarrow g_{1\mathbf{k}} \hat{\mathbf{j}} $, while the second component is proportional to $\mathbf{k} \sigma_{0}$. Because of this, one cannot rewrite $\mathbf{J}(\mathbf{k})$ simply in terms of the transport relaxation time, $\tau_{tr}$, as it can be done in quasiclassical calculations, where $\hat{\mathbf{j}}\to\hat{\mathbf{j}}\, \tau_{tr}/\tau$.

\begin{figure}[t]
	\centering
	\includegraphics[width=1.0\columnwidth]{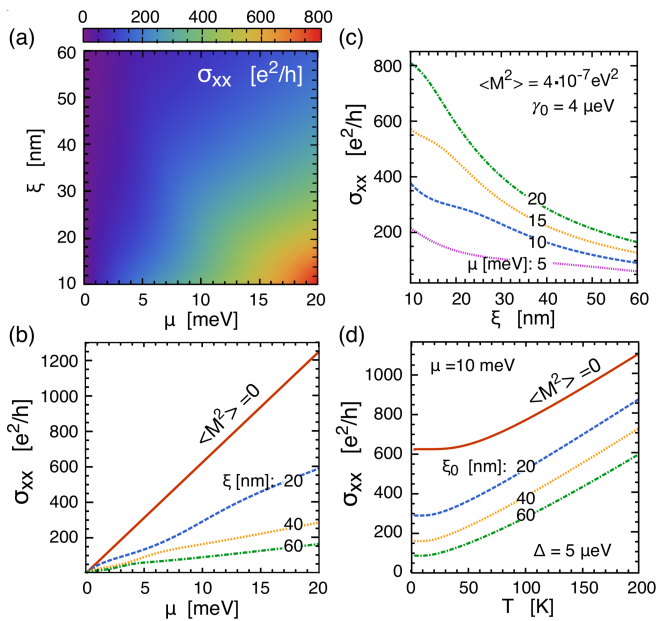}
	\caption{Zero-temperature electrical conductivity as a function of the chemical potential $\mu$ and  correlation length $\xi$ (a), and the corresponding cross-sections for constant values of $\xi$ (b) and constant values of $\mu$ (c). Conductivity in the limit of zero magnetization fluctuations, $<M^2>=0$, is also shown in (b). This conductivity is determined by the relaxation rate $\gamma_0$.  The temperature dependence of the conductivity is shown in (d). Other parameters as in Fig.1. }
	\label{fig2}
\end{figure}

Taking into account the explicit form of the current density vertex function, one finds the diagonal conductivity in the form
\begin{eqnarray}
\label{59}
\sigma_{xx}=-\frac{e^{2} v^{2}}{2\pi \hbar } {\rm Tr}\int \frac{d^2{\bf k}}{(2\pi )^2}
\int d\varepsilon \, f'(\varepsilon )
\left\{ k_{x} g_{0 \mathbf{k}} G^R_{\bf k}(\varepsilon )\, \sigma _x\, G^A_{\bf k}(\varepsilon )\right.\nonumber\\
\left.+ g_{1 \mathbf{k}} \sigma_{x}
 G^R_{\bf k}(\varepsilon )\, \sigma _x\, G^A_{\bf k}(\varepsilon )\right\}. \hspace{1cm}
\end{eqnarray}
Upon integrating the above equation over the wavevector $\mathbf{k}$ one can rewrite it as
\begin{eqnarray}
\sigma_{xx} = - \frac{e^{2}}{8 \pi \hbar} \int d \varepsilon f'(\varepsilon) \left[ \varepsilon^{2} \frac{g_{1 k_1}}{v k_{1} \gamma_{1}} + \varepsilon \frac{k_{1}}{\gamma_{1}} g_{0 k_{1}} \right] \Theta [ \varepsilon ]\nonumber\\
 - \frac{e^{2}}{8 \pi \hbar} \int d \varepsilon f'(\varepsilon) \left[ \varepsilon^{2} \frac{g_{1 k_1}}{v k_{1} \gamma_{2}} + \varepsilon \frac{k_{1}}{\gamma_{2}} g_{0 k_{1}} \right] \Theta [ -\varepsilon ],\hspace{0.5cm}
 \end{eqnarray}
 where $\Theta[\pm \varepsilon]$ is the Heaviside step function.
 Taking into account the relation
 \begin{eqnarray}
 \left[ \varepsilon^{2} \frac{g_{1 k_1}}{v k_{1} \gamma_{1}} + \varepsilon \frac{k_{1}}{\gamma_{1}} g_{0 k_{1}} \right]_{\varepsilon > 0}  =  \left[ \varepsilon^{2} \frac{g_{1 k_1}}{v k_{1} \gamma_{2}} + \varepsilon \frac{k_{1}}{\gamma_{2}} g_{0 k_{1}} \right]_{\varepsilon < 0} =  \frac{v k_{1}}{\gamma \kappa},\nonumber\\
\end{eqnarray}
one can write the conductivity in the following simple final form
\begin{eqnarray}
\label{sig_xx_Final}
\sigma_{xx} = - \frac{e^{2} }{h } \int d\varepsilon f'(\varepsilon) \frac{|\varepsilon|}{4 \gamma \kappa},
\end{eqnarray}
where $f'(\varepsilon)=\partial f/\partial \varepsilon$.
Note that according to Eq.(11) and Eq.(20), both the  relaxation rate, $\gamma$, and the parameter $\kappa$ depend on energy.
%

{\it Results and discussions:} The formula (\ref{sig_xx_Final}) is the starting point for further analysis and discussion of numerical results.
Figure~\ref{fig2} presents the longitudinal electric conductivity plotted for selected  values of the system's parameters. Figure~\ref{fig2}(a) shows the zero-temperature conductivity as a function of the chemical potential, $\mu$, and the correlation length, $\xi$.
The area in the plane ($\xi ,\mu$), where the conductivity reaches large values  is shifted towards smaller $\xi$ and larger $\mu$ in comparison with the area where the  scattering rate is large (compare Fig.~\ref{fig1} and Fig.~\ref{fig2}). This is due to a significant impact of the vertex corrections on the electrical conductivity.
The corresponding cross-sections for constant values of $\xi$ and constant values of $\mu$  are shown in Fig.~\ref{fig2}(b) and Fig.~\ref{fig2}(c), respectively. Note, in Fig.~\ref{fig2}(b) we also show the conductivity  in the absence of the magnetization fluctuations, $<M^2>=0$, where the conductivity  is determined by the scattering rate $\gamma_0$, and increases linearly with the Fermi energy. As follows from the figures, scattering on magnetic fluctuations reduces the conductivity, and for the parameters assumed in Fig.~\ref{fig2} the conductivity for a constant value of $<M^2>$ decreases with increasing correlation length $\xi$.

In Fig.~\ref{fig2}(d) we show the temperature dependence of the conductivity in the absence of electron scattering by phonons.
Interestingly, the conductivity increases with increasing  temperature, also in the absence of the magnetization fluctuations. In the latter case the increase is due to the Fermi distribution and nonconservation of the number of particles in the system when the chemical potential is fixed while the temperature is varied. In the case under consideration, the number of electrons increases with increasing temperature, which gives rise to an increase in  the conductivity.  In the presence of magnetization fluctuations, there is an additional contribution to the temperature dependence of the conductivity, which follows from the temperature dependence of the correlation length described earlier.
%

{\it Summary:} We have studied the impact of correlated fluctuations of magnetization (exchange field) on the transport properties of surface 2D electrons in topological insulators. The fluctuations have been described by their amplitude $\sqrt{<M^2>}$ and correlation length $\xi$. We have also taken into account the reduction of the  correlation length with increasing temperature. It is also worth to note that the description is based on a perturbative approach, so the amplitude of the fluctuations as well as the appropriate correlation length cannot be arbitrarily large.
To infer the electrical conductivity we  determined at first the relaxation rate due to scattering on the magnetization fluctuations, and then the appropriate current density vertex function. The latter turned out to have a significant influence on the conductivity. The conductivity was calculated assuming additional scattering on other structural defects, as described by the scattering rate $\gamma_0$. In general, the conductivity is remarkably reduced by scattering on the magnetization fluctuations. The temperature dependence of the magnetization fluctuations  was also included by a phenomenological formula, and was shown to have a remarkable impact on the transport properties. Thus, the temperature dependence of the conductivity follows not only from the Fermi distribution, but also from the temperature dependence of the correlation length.

\begin{acknowledgments}
This work is supported by the National Science Center in Poland under Grant No. DEC-2017/27/B/ST3/02881. A. Dyrda\l{}  acknowledges the support of German Research Foundation (DFG) through SFB 726
\end{acknowledgments}

\end{document}